\documentclass[amsmath,amssymb,prl,aps,showpacs,twocolumn,superscriptaddress,floatfix,footinbib]{revtex4-1}

\usepackage{graphicx,fancybox}
\usepackage{amsmath,amssymb,mathrsfs,slashed}
\usepackage[title]{appendix}
\usepackage[colorlinks=true, pdfstartview=FitV, linkcolor=red, citecolor=blue, urlcolor=blue]{hyperref}
\usepackage{color}
\usepackage{soul}
\usepackage{array}
\usepackage{url}
\usepackage{ulem}
\usepackage{bm}
\usepackage{dcolumn}
\usepackage{bm}
\usepackage{braket,xfrac,lineno}
\usepackage[utf8]{inputenc}
\usepackage{tikz}
\usepackage{tikz-feynman}
\tikzfeynmanset{
    every diagram={thick},
}

\begin{document}
\newcommand \etc {{\it etc.} }
\newcommand \tie {{\it i.e.}}
\newcommand \ie {{\it i.e.} }
\newcommand \f {\not\!}
\newcommand \wh {\widehat}
\newcommand \hq {\hat{q}}
\newcommand \hk {\hat{k}}
\newcommand \hw {\hat{w}}
\newcommand \nf {\tilde{n}}
\newcommand \kq {E_{k-q}}
\newcommand \qk {E_{q-k}}
\newcommand \pk {E_{p-k}}
\newcommand \uqk {\widehat{(q-k)}}
\newcommand \np {\tilde{n}^{\prime}}
\newcommand \kd  {\delta}
\newcommand \ra  {\rightarrow}
\newcommand \ev {\delta \Gamma}
\newcommand \po {$\Pi^{1}$\,}
\newcommand \pt {$\Pi^{2}$\,}
\newcommand \w  {\omega}
\newcommand \fw {{\bf w}}
\newcommand \fp {{\bf p}}
\newcommand \fn {{\bf n}}
\newcommand \fpo {{\bf p_1}}
\newcommand \fpt {{\bf p_2}}
\newcommand \fk {{\bf k}}
\newcommand \fq {{\bf q}}
\newcommand \fx {{\bf x}}
\newcommand \fy {{\bf y}}
\newcommand \fl {{\bf l}}
\newcommand \fgma {{\bf \gamma}}
\newcommand \h {\theta}
\newcommand \im {\Rightarrow}
\newcommand \vk {\vec{k}}
\newcommand \vl {\vec{l}}
\newcommand \vq {\vec{q}}
\newcommand \vw {\vec{w}}
\newcommand \vx {\vec{x}}
\newcommand \vp {\vec{p}}
\newcommand \vecr {\vec{r}}
\newcommand \lra {\leftrightarrow}
\newcommand \mat {{\mathcal M}}
\newcommand \mal {{\mathcal L}}
\newcommand \mps {\mu}
\newcommand \mas {\bar{\mu}}
\newcommand \g {\gamma}
\newcommand \ro {\rho}
\newcommand \si {\sigma}
\newcommand \e {\epsilon}
\newcommand \ve {\varepsilon}
\newcommand \onp {(1/2 - \tilde{n}(k))^{\prime}}
\newcommand \p {^{\prime}}
\newcommand \N {{\mathcal N}}
\newcommand \Sc {{\mathcal S}}
\newcommand \x {\cdot}
\newcommand \hf {\frac{1}{2}}
\newcommand \A {\alpha}
\newcommand \B {\beta}
\newcommand \tbe { \log \left( \frac{2}{\e} \right) }
\newcommand \pr {\hspace{1.5cm}}
\newcommand \lc {\langle}
\newcommand \rc {\rangle}
\newcommand \prt {\partial}
\newcommand \D {\Delta}
\newcommand \sg {\sigma}
\newcommand \eo {\epsilon_0}
\newcommand \nt {\noindent}
\newcommand \T {\tilde}
\newcommand \al {\alpha}
\newcommand \Ow {\Omega}
\newcommand \dsc {\mbox{Disc}}
\newcommand \Op {{\mathcal O}}
\newcommand \Ro {\mathcal{R}}
\newcommand \ad {a^{\dag}}
\newcommand \bd {b^{\dag}}
\newcommand \ua {\uparrow}
\newcommand \da {\downarrow}
\newcommand {\llb} { \left[ \frac{\mbox{}}{\mbox{}} \right.}
\newcommand {\lrb} { \left. \frac{\mbox{}}{\mbox{}} \right] }
\newcommand \mt {\mathcal{T}}
\newcommand \mb {\mathcal{B}}
\newcommand \mh {\mathcal{H}}
\newcommand \mhp {\mathcal{H'}}
\newcommand \ml {\mathcal{L}}
\newcommand \gmn {g^{\mu \nu}}
\newcommand \gmr {g^{\mu \ro}}
\newcommand \gnr {g^{\nu \ro}}
\newcommand \af {\frac{1}{2}}
\newcommand \ini {\infty}
\newcommand \res {\mbox{Res.}}
\newcommand \kD {\Delta}
\newcommand \ma {\mathcal{A}}
\newcommand \bvec{\left( \begin{array}{c} }
\newcommand \evec{\end{array} \right)}
\newcommand \bmat{\left( \begin{array}}
\newcommand \emat{\end{array} \right)}
\newcommand \tr {\mbox{{\bf Tr}}}
\newcommand \eg {{\it e.g.}}
\newcommand \bea{\begin{eqnarray} }
\newcommand \eea{\end{eqnarray} } 
\newcommand \nn {\nonumber}
\newcommand {\be} {\begin{equation}}
\newcommand {\ee} {\end{equation}}
\newcommand {\epem} {$e^+ e^-$}
\newcommand {\mbx} {\mbox{}}
\newcommand {\gol} {\bullet}
\newcommand {\gev} {\mbox{GeV}}
\newcommand {\ata} {& \times &}
\newcommand {\psibar} {\bar{\psi}}

\definecolor{hu-berlin-blue}{RGB}{0,65,137} 
\definecolor{hu-berlin-green}{RGB}{150,190,20} 
\definecolor{hu-berlin-grey}{RGB}{169,169,169}
\definecolor{hu-berlin-brown}{RGB}{82,79,60}
\definecolor{hu-berlin-red}{RGB}{180,0,0}

\newcommand{\hublue}{\color{hu-berlin-blue}}
\newcommand{\hugreen}{\color{hu-berlin-green}}
\newcommand{\hugrey}{\color{hu-berlin-grey}}
\newcommand{\hubrown}{\color{hu-berlin-brown}}
\newcommand{\hured}{\color{hu-berlin-red}}

\newcommand{\JHW}[1]{\textcolor{green}{#1}}
\newcommand{\AM}[1]{\textcolor{red}{#1}}
\newcommand{\AK}[1]{{\textcolor{red}{#1}}}
\newcommand{\IS}[1]{{\textcolor{blue}{#1}}}

\title{Fluctuation-Dissipation Relation for Hard Partons in a Gluonic Plasma}

\author{Amit~Kumar} 
\affiliation{Department of Physics, University of Regina, Regina, Saskatchewan S4S 0A2, Canada}

\author{Abhijit~Majumder}
\affiliation{Department of Physics and Astronomy, Wayne State University, Detroit, MI 48201, USA}

\author{Ismail Soudi}
\affiliation{Faculty of Physics, Bielefeld University, Bielefeld D-33615, Germany}

\author{Johannes~Heinrich~Weber} 
\affiliation{Institut f\"ur Kernphysik, Technische Universit\"at Darmstadt, Schlossgartenstra\ss e 2, D-64289 Darmstadt, Germany}
\affiliation{K{\"a}the-Kollwitz-Gymnasium, Villenstra{\ss}e 1, D-67433 Neustadt/Wstr., Germany}

\date{\today}

\begin{abstract}
We derive a fluctuation dissipation relation connecting the drag and diffusion jet transport coefficients for an energetic light quark traversing a non-perturbative thermalized gluon plasma. The hard quark is taken to be close to on-shell, with an energy scale parametrically larger than the medium temperature.
We introduce a general complex-valued function for each transport coefficient. Evaluating these in the deep Euclidean momentum region enables their expression in terms of local operators. 
Using contour-integration techniques, we relate these local operators, after vacuum subtraction, to the physical transport coefficients that arise along a branch cut, close to light-like dispersion. 
The derived relation relates the longitudinal drag coefficient to the longitudinal and transverse diffusion coefficients, and the thermal gluon condensate.
\end{abstract}

\maketitle



To the best of our knowledge, the Quark Gluon Plasma (QGP) formed in heavy-ion collisions at the Relativistic Heavy-Ion Collider (RHIC) and at the Large Hadron Collider (LHC), at temperatures $T\!\gtrsim\! \Lambda_{\rm QCD}$, is strongly coupled~\cite{Romatschke:2007mq,Song:2010mg,JETSCAPE:2020mzn}. Hard partons (with energy $E \gg T$), within high transverse momentum jets, are expected to be weakly coupled with the QGP, due to their large energy scale~\cite{Baier:1994bd,Baier:1996kr}. This allows for the application of perturbation theory~\cite{Majumder:2010qh} to the scattering and radiation process, while the color field of the medium, off which the hard parton scatters, may indeed be non-perturbative~\cite{Wang:2001ifa,Sirimanna:2021sqx}. 

In the above limit, for a hard quark projectile, one may decompose the single scattering process into the hard quark interacting via single gluon exchange with a strongly coupled medium~\cite{Majumder:2012sh,Kumar:2020wvb}. However, in most current, successful, jet modification calculations~\cite{JETSCAPE:2022jer,JETSCAPE:2022hcb,JETSCAPE:2023hqn,JETSCAPE:2024nkj}, the exchanged gluon is assumed to be obtained from a weakly coupled QGP~\cite{Schenke:2009gb,He:2015pra}, even though the bulk dynamics of the QGP may be strongly coupled. 
This is often referred to as the {\it recoil-hole} scheme~\cite{Zapp:2008gi}.

The choice of a specific model for the medium (in the vicinity of the jet), in this case a weakly coupled thermal plasma~\cite{Braaten:1989mz,Frenkel:1989br}, immediately implies the existence of an order-by-order calculable differential scattering rate $\sfrac{d^3 \Gamma}{d^3 k}$~\cite{Caron-Huot:2008zna}, where $k$ represents the exchanged momentum between the medium and the hard parton. This in turn implies the existence of fluctuation-dissipation relations between the first non-zero moments of this distribution
(generated by the same perturbative kernel):
\begin{align}
    \mbox{}\!\!\!\!\hat{q} \!&=\!\!\! \int\!\! d^3 \!k  \pmb{k_\perp}^{\!\!2} \frac{d^3 \Gamma}{d^3 k} ,\, 
    \hat{e} \!=\!\!\! \int\!\! d^3 \!k  k^- \frac{d^3 \Gamma}{d^3 k} , \, 
    \hat{e}_2 \!=\!\!\!\int\!\! d^3 \!k  (k^-)^2 \frac{d^3 \Gamma}{d^3 k}.\!
    \label{eq:transport_coefficients}
\end{align}
In the above equation, $\hat{q}$ is the transverse momentum diffusion coefficient~\cite{Baier:2002tc}, $\hat{e}$ is the longitudinal light-cone drag coefficient, where $k^-= \sfrac{(k^0 - k^3)}{\sqrt{2}}$, [assuming the hard jet travels in the ($-z$) direction], and $\hat{e}_2$~\footnote{Instead, one may define $\hat{e}_2\! =\! \langle (k^-)^2 \rangle - \langle k^- \rangle^2$, we show that the $2^{\rm nd}$ term is suppressed at high jet parton energy.} represents the diffusion in longitudinal light-cone momentum~\cite{Majumder:2008zg}. 

In spite of the advantages that the assumption of a weakly coupled medium affords, and its success in comparison with experimental data~\cite{Cao:2024pxc}, detailed Bayesian analysis using a wide range of observables yields tensions in the extracted values of $\hat{q}$~\cite{JETSCAPE:2024cqe}. The main issue is the shape of $\hat{q}$ as a function of the local temperature $T$, as seen in plots of $\hat{q}/T^3$. Calculated within a weakly interacting quasi-particle medium, $\hat{q}/T^3$ contains an unphysical rise, as $T$ drops down to the transition temperature $T_C$, below which the theory is inapplicable. Based on simple arguments (and calculations in lattice QCD~\cite{HotQCD:2014kol}), the medium, along with the number of scattering centers, should diminish at $T \ll T_C$, leading to a positive slope of $\hat{q}/T^3$ at low temperatures, while the running of the coupling at very high temperatures should lead to a decreasing $\hat{q}/T^3$ at $T\gg T_C$ (negative slope). One thus expects a plateau like behavior for $\hat{q}/T^3$ with a maximum close to $T_C$. This behavior is indeed reproduced by calculations of $\hat{q}/T^3$ in lattice QCD~\cite{Kumar:2020wvb}. In fact, modifications of the scattering rate that yield a plateau like form for $\hat{q}/T^3$~\cite{Ren:2026xfn} yield improved comparisons with data~\cite{Datta:2025gql}.  

Beyond $\hat{q}$, little is known about the non-perturbative aspects of the other coefficients $\hat{e}$ and $\hat{e}_2$, which may indeed introduce significant modifications in comparisons of jet modification calculations with data~\cite{Qin:2009gw}. While $\hat{e}_2$ can be calculated in lattice QCD using methods similar to those for $\hat{q}$ (both represent fluctuations), there is no way to calculate $\hat{e}$ (which represents a time dependent process) on the lattice. The only means to obtain this transport coefficient would be using a fluctuation dissipation relation, where the remaining coefficients are known. 

The goal of this Letter is to demonstrate that a relation between the three leading transport coefficients can be derived non-perturbatively. In fact, the only required assumption is that the medium is a thermalized, homogeneous, isotropic, $SU(3)$ plasma, invariant under parity and time reversal. The derived fluctuation dissipation relation holds for both strongly coupled and weakly coupled media, clearly highlighting the change as one transitions from strong to weak coupling (the use of this relation to obtain $\hat{e}$ will be presented in an upcoming paper~\cite{KMSW:2}). Beyond its applicability to elucidate jet modification calculations, a fluctuation dissipation theorem has never been derived before without assuming a model for the scattering kernel and the medium~\cite{Ghiglieri:2015ala}, as in Eq.~\eqref{eq:transport_coefficients}.


Jet transport coefficients governing partonic drag and diffusion have been derived, in the limit that the interaction of the hard parton is perturbative, but the medium may or may not be perturbative~\cite{Majumder:2012sh, Kumar:2020wvb, Benzke:2012sz, Panero:2013pla, Laine:2013apa, Antonov:2007sh, Abir:2015hta, Abir:2014sxa, Idilbi:2008vm, Liu:2006ug,Lin:2006au,Avramis:2006ip, Bjorken:1982tu, THOMA1991128, Thoma:1990fm, Braaten:1991we, Peigne:2008nd, Peng:2024zvf}. In operator product expansion, the coefficients are expressed in terms of nonperturbative two-point gluonic correlation functions, which encode the interaction of energetic partons with the medium. 

\begin{figure}[h!]
    \begin{center}
    \includegraphics[width=0.4\textwidth]{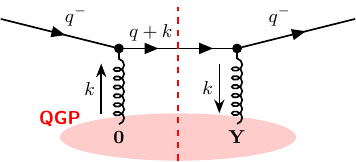}
    \end{center}
  \caption{A forward scattering diagram for the hard quark undergoing a single scattering 
  off the gluon field in the plasma. The vertical dashed line represents the cut-line.}
  \label{fig:ForwardScatteringDiagram}
\end{figure}

We consider a three-dimensional (3D) box filled with a static plasma, with equal spatial extents $L_{x}=L_{y}=L_{z}=L$. 
We study a leading-order process in which an energetic on-shell quark, with four momentum $q^{\mu}= [q^{+}, q^{-}, \pmb{q_{\perp}}=\pmb{0_{\perp}} ] $, propagates in the negative $z$-direction. 
The incoming quark undergoes a single scattering via the exchange of a gluon with the plasma, and subsequently exits the medium with momentum $q^{\mu}+k^{\mu}$. 
In this setup, the incoming quark carries a large light-cone momentum component $q^{-} \gg \Lambda_{\rm QCD}$, which sets the hard scale of the problem. 
The on-shell condition requires the corresponding $(+)$ light-cone component $q^{+}\approx 0$. 
The exchanged gluon carries momentum $k^{\mu}$, whose components lie in the Glauber kinematic region, i.e., the transverse momentum is large compared to its $(+)$ and $(-)$ light-cone components. 
If the transverse momentum scales as $\pmb{k_{\perp}} \sim \lambda q^-$, the on-shell condition for the outgoing quark, $(q+k)^2=0$, constrains the $(+)$ light-cone momentum to scale as $k^+ \sim \lambda^2 q^-$. 
The $(-)$ light-cone component $k^-$ may scale either as $k^+$ or as $|\pmb{k_{\perp}}|$, depending on the kinematics.  
This implies that the $\lambda$ power counting for the Glauber gluon is $[k^{+}, k^{-}, \pmb{k_{\perp}}] \sim [\lambda^2, \lambda, \lambda]q^-$, where $\lambda$ is a small dimensionless parameter such that $\lambda^{2} \ll \lambda \ll 1$.

A representative forward-scattering diagram is shown in Fig.~\ref{fig:ForwardScatteringDiagram}, illustrating the exchange of a gluon with momentum $k^{\mu}$ between the incoming quark and the plasma. The expectation of an observable $O$ 
is given as
\begin{align}
  \mbox{}\!\!\!\!\langle \hat{\mathcal{O}} \rangle     
  & = \frac{1}{t} \sum_{k} \hat{\mathcal{O} } \text{ } \frac{d^4\Gamma}{dk^4}  
  = \frac{g^2}{ 2N_{c}2E_{q}}\int d^4 Y  \frac{d^{4}k}{(2\pi)^4} e^{ikY}  
  \hat{ \mathcal{O} } 
    \label{eq:general_relation_operator_O} \\
  &\times  2\pi \delta[(q+k)^2] 
 \bra{M}  \mathrm{Tr} \left[ \slashed{q}   \slashed{A}\left( 0 \right) 
  (\slashed{q} + \slashed{k})  \slashed{A}\left( Y \right)  \right] \ket{M}, 
\nonumber
\end{align}
where $t=L/c$ is the time spent by the quark inside the box,  $\frac{d^4\Gamma}{dk^4} $ is the differential scattering rate (performing incoming quark's spin summation and color averaging), $g=\sqrt{4\pi \alpha_s}$ is the strong coupling constant at the vertex of the hard quark and the exchanged gluon, $E_q$ is the incoming quark energy, and $A^{\mu}(x)=t^{a}A^{\mu,a}(x)$ is the gluon vector potential with $t^{a}$ being the Gell-Mann matrices for SU(3) color algebra. 
The $\bra{M} \ldots \ket{M}$ represents the thermal ensemble average over all possible initial states of the medium.
The choice of the operator $ \hat{ \mathcal{O}  }  $ depends on the transport coefficient in question. 
Thus, $\hat{\mathcal{O}} = \pmb{k_\perp}^{\!\!2} $ for $\hat{q}$, $ k^-$ for $\hat{e}$ and $(k^-)^2$  for $\hat{e}_2$.

Substituting the operator expressions for each transport coefficient, we obtain:
\begin{eqnarray}
 \hat{q} \! & =& \!  c_{0}\! \int \frac{d^{4}Y d^4 k}{(2\pi)^4}  
e^{ik \cdot Y } 2\pi \delta \left( k^+ - \frac{k_\perp^2}{2(q^- + k^ -) } \right)  \nonumber \\ 
 &\times&  \bra{M}  \mathrm{Tr} 
 \!\!\left[ {F_{\perp}}^{\!\!\alpha +}(0) {F_{\perp}}^{\!\!+}_{,\alpha}(Y) \right]\!\! \ket{M} ,
 \label{eq:expectation_k2_perp} \\
\hat{e}_2 \! &=& \!  c_{0}\! \int \frac{d^{4}Y d^4 k}{(2\pi)^4}  
e^{ik \cdot Y } 2\pi \delta \left( k^+ - \frac{k_\perp^2}{2(q^- + k^ -) } \right)  \nonumber \\ 
 & \times& 
  \Biggl[ \bra{M}  \mathrm{Tr} \left[ F^{-+}(0) F^{-+}(Y) \right] \ket{M}
  \label{eq:expectation_k2_minus} \\
  &+&   \!\!\! \frac{q^-}{2}  \sum_{n=2}^{\infty} \left(\frac{-i\partial^-}{q^-} \right)^n\bra{M}  
 \mathrm{Tr} \left\{  
  A(0) \cdot \partial \cdot A (Y)\right\} \ket{M} \Biggr] ,
  \nonumber \\
  \hat{e} \! &=& \! c_{0}\! \int \frac{d^{4}Y d^4 k}{(2\pi)^4}  
e^{ik \cdot Y } 2\pi \delta \left( k^+ - \frac{k_\perp^2}{2(q^- + k^ -) } \right)  \nonumber \\ 
 & \times & 
\Biggl[  \text{ } \bra{M}  \mathrm{Tr} \left[  [-iF^{-+}(0)]  A^{+}(Y) \right] \ket{M}
\label{eq:expectation_k_minus}  \\
&-& \frac{1}{2} \sum_{n=1}^\infty \left( \frac{-i \partial^-}{ q^-}  \right)^n \bra{M}  
  \mathrm{Tr} \left\{   
  A(0) \cdot \partial \cdot A (Y)\right\} \ket{M} \Biggr]. \nonumber
\end{eqnarray}
In the above equations, $c_{0}=\sfrac{8\pi \alpha_s}{ (\sqrt{2}N_{c})}$, 
In Eqs.~(\ref{eq:expectation_k2_minus},\ref{eq:expectation_k_minus}), $A(0) \cdot \partial \cdot A (Y)$ is a shorthand for the factor  
\[A^+(0) i \prt_{\alpha} A_\bot^\alpha(y) + A_\bot^\alpha(0) i \prt_\alpha A^+ (y) - A_\alpha(0) i \prt^+ A_\bot^\alpha(y),\] 
where $A_\bot^\alpha \neq 0$, only if $\alpha = 1,2$. Similarly, the subscript in $F_\perp^{\alpha \beta}$ indicates that either $\alpha$ or $\beta$ is limited to 1,2.

We point out that Eq.~\eqref{eq:expectation_k2_perp} is independent of gauge and exact up to terms suppressed by powers of coupling. In the 2$^{\rm nd}$ lines of  Eq.~\eqref{eq:expectation_k2_minus} and \eqref{eq:expectation_k_minus}, we chose $A^-=0$ gauge and have replaced $\prt^- A^+ $ as $F^{-+}$. 
No modification based on $A^-\!\!=0$ gauge has yet been applied to the $A(0) \cdot \partial \cdot A (Y)$ factor; this will be carried out after cancellations between $\hat{e}$ and $\hat{e}_2$. 
Although the formal expressions for $\hat{e}$ and $\hat{e}_2$ vary with choice of gauge, the actual values of (or derived relations between) the coefficients are gauge invariant. Derivations of $\hat{e},\hat{e}_2$ in other gauges and evaluation on a lattice will be presented in an upcoming effort~\cite{KMSW:2}.

In order to make the sums in $\hat{e}$ and $\hat{e}_2$ identical, we extract the $n=1$ term in $\hat{e}$ and reexpress the $\prt^-$ derivative, using translational invariance of the matrix element: 
\begin{eqnarray}
&&\!\!\!\!\int d^4 Y e^{ikY} {i\prt ^-}\bra{M} A(0) \cdot \prt \cdot A(Y) \ket{M} = \int d^4 Y e^{ik\cdot Y} \nonumber \\
&& \!\!\!\!\left[ \bra{M} \!\!\mathrm{Tr} \left[ {F_{\!\bot}}_\alpha^{\,\,,+} (0) {F_{\!\perp}\!\!}^{- \alpha} (Y) +  
{F_{\!\perp}\!\!}^{- \alpha} (0)  {F_{\!\perp}}_\alpha^{\,\,,+} (Y) \right] \!\!\ket{M} \right. \nonumber \\
%
%
%
%
&& \!\!\!\!\left. -i\bra{M} \!\! \frac{A_{\alpha}(0)\nabla^2_{\!\perp}F^{-\alpha}_{\perp}(Y)}{2q^-} \! + \!\!\frac{F^{-}_{\perp,\alpha}(0)\partial^+ F^{-\alpha}_{\perp}(Y)}{q^-}\!\!\ket{M}\!\right]\!. 
\label{eq:AdA_expansion}
\end{eqnarray}
In the last line above, we have used the overall delta function to convert $k^+ = \sfrac{k_\perp^2}{2(q^- + k^-)}$ and express $k^{+}=\sfrac{[k^2_{\perp} - 2k^-k^+]}{(2q^-})$.

Thus, the only difference in the structure of the expectation of the exchanged momentum is the combination of field strength tensors, derivatives, and vector potentials that arise in each expression. We can thus express all three expectations using the generic form: 
\begin{eqnarray}
    \langle O \rangle &=& c_{0}\int d^4 Y  \frac{d^{4}k}{(2\pi)^4}  \text{ } e^{ik \cdot Y} 2\pi \delta \left( k^+ - \frac{k_\perp^2}{2(q^- + k^ -) } \right) \nonumber \\
    & & \times \bra{M}  \mathrm{Tr} \left[ \mathcal{O}_{1}(0) \mathcal{O}_{2}(Y)   \right] \ket{M} .
\end{eqnarray}
The non-local operator product  $\mathcal{O}_{1}(0) \mathcal{O}_{2}(Y)$ is constructed from in-medium gluon field operators, and possibly their derivatives. At this point, there is no obvious fluctuation dissipation relation among the drag and diffusion coefficients.


Analytically continuing the light-cone momentum $q^+$, we define a complex-valued function $\mathcal{C}(q^{+}) $ as follows 
\begin{align}
     \mathcal{C}(q^{+}) &= c_{0} \int d^4Y  \frac{d^{4}k}{(2\pi)^4}  \text{ } e^{ik Y} \left[ 2(q^{-}+k^{-}) i \right]  \nonumber \\
     & \times \frac{\bra{M}  \mathrm{Tr} \left[ \mathcal{O}_{1}(0) \mathcal{O}_{2}(Y)   \right] \ket{M}}{(q+k)^2+i\epsilon} ,
     \label{eq:complex-function}
\end{align}
where $q^{+}$ is treated as a complex-number, and $q^-$ is held fixed. 
In the vicinity of $q^{+}\approx 0$, we obtain
\begin{align}
    &\mathrm{Disc}\left[ \frac{1}{(q+k)^2 +i\epsilon}\right]_{ {\rm at} \text{ } q^{+}\approx 0} = -2\pi i \delta[(q+k)^2]  \nonumber \\
    & = \frac{-2\pi i}{2(q^- + k^-)} \delta \left( k^{+} - \frac{\pmb{k}^2_{\perp}}{2 (q^- + k^-) } \right).
\end{align}
Thus, ${\rm Disc} \left[ \mathcal{C}(q^{+}) \right]_{ {\rm at} \text{ } q^{+}\approx 0} = \langle O \rangle$.
The physical transport coefficients $\langle O \rangle$ can be obtained from the discontinuity of the analytically continued (non-physical) function $C(q^+)$ across its branch cut near $q^+\approx 0$. Assuming a medium with temperature $T$, the branch cut is at most a factor of approximately $T$ on either side of $q^+=0$.

The complex-valued function $\mathcal{C}(q^{+})$ can be evaluated at $q^{+}=-M_{\infty}$, where $M_{\infty} \gg q^{-}$ (deep Euclidean region), and expressed in terms of local operators as,
\begin{eqnarray}
    & &\lim_{M_{\infty} \rightarrow \infty}  \mathcal{C}(q^{+}=-M_{\infty})  \nonumber \\
    &=&  \lim_{M_{\infty} \rightarrow \infty} c_{0} \int   \frac{d^4Y d^{4}k}{(2\pi)^4}  \text{ } e^{ik Y} \left[ 2(q^{-}+k^-)i \right] \nonumber \\
    && \times \frac{\bra{M}  \mathrm{Tr} \left[ \mathcal{O}_{1}(0)  \mathcal{O}_{2}(Y)   \right] \ket{M}}{\left[ 2(q^{-}+k^{-})(-M_{\infty}+k^{+})-\pmb{k}^2_{\perp}\right]}  \nonumber \\
    & =& \lim_{M_{\infty} \rightarrow \infty} c_{0}  \left[ \frac{-i }{M_{\infty}}  \right] \int d^4Y  \frac{d^{4}k}{(2\pi)^4}  \text{ }  e^{ik Y}  \nonumber \\
    && \times \bra{M}  \mathrm{Tr} \left[ \mathcal{O}_{1}(0) \mathcal{O}_{2}(Y)   \right] \ket{M} \nonumber \\
    &=&  \lim_{M_{\infty} \rightarrow \infty} c_{0}  \left[ \frac{-i }{M_{\infty}}  \right]   \bra{M}  \mathrm{Tr} \left[ \mathcal{O}_{1}(0)  \mathcal{O}_{2}(0)   \right] \ket{M}. 
\end{eqnarray}
We now consider the contour integral around a circle centered at $q^{+}=-M_{\infty}$ (where $ M_{\infty} \rightarrow \infty$), given as
\begin{eqnarray}
    I_{1} &=& \oint \frac{dq^+}{2\pi i} \frac{\mathcal{C}(q^+)}{(q^{+}+M_{\infty})} = \mathcal{C} (q^{+}=-M_{\infty}) ,
\label{eq:contour_integral_def}
\end{eqnarray}
where the contour is oriented in a counterclockwise direction. The contour can be deformed along the branch cut $q^{+} \in (-T_{1},\infty)$, allowing the integral to be written as
\begin{eqnarray}
    && \lim_{M_{\infty} \rightarrow \infty} \oint \frac{dq^+}{2\pi i} \frac{\mathcal{C}(q^+)}{(q^{+}+M_{\infty})} \nonumber \\
    &=& \int^{T_{2}}_{-T_{1}} \frac{dq^{+}}{2\pi i} \frac{{\rm Disc}[\mathcal{C}(q^+)]}{(q^{+}+M_{\infty})}  +  \int^{\infty}_{0} \frac{dq^{+}}{2\pi i} \frac{{\rm Disc}[\mathcal{C}(q^+)]}{(q^{+}+M_{\infty})} \nonumber \\
    &=&\!\! \lim_{M_{\infty} \rightarrow \infty} \!\!\frac{\langle O\rangle \text{ } (T_{1}+T_{2})}{M_{\infty} \text{ } 2\pi i} +  \int^{\infty}_{0} \frac{dq^{+}}{2\pi i} \frac{{\rm Disc}[\mathcal{C}(q^+)]}{(q^{+}+M_{\infty})} .
    \label{eq:discont_T1T2_O}
\end{eqnarray}
In Eq.~\eqref{eq:discont_T1T2_O}, $T_1 +T_2 \equiv \delta_T T $ represents a width of the thermal discontinuity of $\mathcal{C}(q^+)$ along the $q^+$ real axis (Note that we have expressly picked $M_\infty \gg T_1$). 
The second integral represents the vacuum discontinuity and encapsulates the contribution from vacuum-like processes. To extract the purely thermal contribution, we subtract the vacuum piece to obtain,
\begin{eqnarray}
  \lim_{M_{\infty} \rightarrow \infty}   \frac{\langle O\rangle \delta_T T }{M_{\infty} \text{ } 2\pi i} &=& \lim_{M_{\infty} \rightarrow \infty} \oint \frac{dq^+}{2\pi i} \frac{\mathcal{C}(q^+)_{\rm Thermal-Vacuum}}{(q^{+}+M_{\infty})} \nonumber \\ 
   =  \lim_{M_{\infty} \rightarrow \infty}&& \!\!\!\!\!\!\!\!\mathcal{C}(q^{+}=-M_{\infty})_{\rm Thermal-Vacuum} .
    \label{eq:O_vacuum_subt_expression}
\end{eqnarray}

The meaning of the vacuum subtraction depends on how the calculation is carried out, e.g., if the local operators are evaluated in lattice QCD, one re-evaluates the expectation for the case of a vacuum lattice, and subtracts from the thermal contribution; if the calculation is done in perturbation theory, one simply does not calculate vacuum like diagrams. 
Simplifying Eq.~\eqref{eq:O_vacuum_subt_expression}, we arrive at the following expression:
\begin{align}
\mbox{}\!\!\!\!\langle O\rangle_{\rm T-V} 
&=  \Tilde{c}_{0} \bra{M}  \mathrm{Tr} \left[ \mathcal{O}_{1}(0)  \mathcal{O}_{2}(0)   \right] \ket{M}_{\rm T-V}.  \label{eq:O_vacuum_subt_expression_final}
\end{align}
where 
we define $\Tilde{c}_{0}=2\pi c_{0}/(\delta_{T} T)$, which contains the jet-medium coupling $\alpha_s$ and width of the thermal discontinuity of $\mathcal{C}(q^+)$ in the vicinity of $q^+ \approx 0$.
%


Going forward, we will not write the $T-V$ subscript and the `Tr' before each operator product, \emph{both should always be assumed to be present}. Once expressed as local operators, the constraints of isotropy, and parity and time reversal invariance can be applied to simplify the operators. 
For the case of $\langle \pmb{k_\perp}^{\!\!2} \rangle$, this yields, 
\begin{eqnarray}
    \hat{q} &=& \Tilde{c}_{0} \bra{M}   \sum_{\alpha=1,2}  F^{\alpha +}_{\!\perp}(0) {F_{\!\perp}}^{\!\!+}_{, \alpha}(0)  \ket{M}  \nonumber \\
&=& \Tilde{c}_{0} \bra{M}   \sum_{j=1,2}\frac{[F^{0j}F^{0j} + F^{3j}F^{3j}] + 2 F^{3j}F^{0j} }{2}\ket{M}\nonumber \\    
 &=&  \Tilde{c}_{0} \bra{M}   [ F^{03}F^{03} + F^{31}F^{31} ] \ket{M},
 \label{eq:k2_perp_fnl}
\end{eqnarray}
where we assume that the medium is isotropic in $x,y$, and $z$ directions, and drop the location $(0)$ in the argument of the operators. 
Note, the antisymmetric operator $F^{31}F^{01}$ in Eq.~\eqref{eq:k2_perp_fnl} is odd under parity as well as time-reversal, therefore, its expectation  $\langle F^{31}F^{01}\rangle$ vanishes. Following similar steps as above, for $\langle (k^-)^2\rangle$, we obtain
\begin{align}
 \mbox{}\!\!\!\!  \hat{e}_2 \! &= \!  \tilde{c}_{0}
 \bra{M} \left\{   (F^{03})^2
 + \frac{q^-}{2} {\!\!\sum\limits_{n=3,5}^{\infty} \!\!   
    \!\left(\!\! \frac{-i\prt^-}{q^-} \!\!\right)^{\!\!n}\!\!\! A\!\cdot\!\prt\!\cdot\! A } \right\}  \ket{M} , \label{eq:e2_hat_local}
\end{align}
where, the $n^{\rm th}$ order derivatives are taken before both vector potentials in the $(A\cdot \prt \cdot A)$ term are evaluated at $Y=0$. In deriving the above equation, all terms with an odd \emph{total} number of derivatives (odd under parity and time reversal) are dropped.

Following similar steps for the drag $\langle k^-\rangle$, we obtain, 
\begin{eqnarray}
\mbox{}\!\!\!\!     \hat{e} \!=\! \tilde{c}_{0}\!  \bra{M} \!\! \left[\! \frac{(F^{03})^2\!\! -\! (F^{13})^2 }{q^-}  
\!- \! \!\!\sum_{n=3,5}^\infty 
 \!\!\!\left( \!\frac{-i\prt^-}{q^-} \!\!\right)^{\!\!n}  \frac{\!\!A\cdot\!\prt\cdot \!A\! }{2} \!   \right]\!\! \ket{M}\!. \!
%
%
%
\label{eq:e-hat_local}
\end{eqnarray}
The terms $\bra{M}A_{\alpha}(0)\nabla^2_{\!\perp}F^{-\alpha}_{\perp}(0) + F^{-}_{\perp,\alpha}(0)\partial^+ F^{-\alpha}_{\perp}(0)\ket{M}$, from the last line of Eq.~\eqref{eq:AdA_expansion} are not present in Eq.~(\ref{eq:e-hat_local}). These are antisymmetric under parity and time reversal, and do not contribute to $\hat{e}$.
Similarities between the expressions for $\hat{e}$ and $\hat{e}_2$ allow us to immediately obtain, 
\begin{align}
 \mbox{} \!\!\!\hat{e} \! &=\! -\frac{\hat{e}_2}{q^-} + \frac{ \tilde{c}_0 \bra{M} 2(F^{03})^2 -  (F^{13})^2 \ket{M}}{q^-}. 
  %
 \label{eq:ehat_related_to_e2hat}
\end{align}
%

Writing $(F^{03})^2 = \sfrac{1}{2} [ (F^{03})^2 + (F^{13})^2 + (F^{03})^2 - (F^{13})^2 ]$, we obtain the fluctuation dissipation relation as, 
\begin{eqnarray}
\mbox{} \!\!\!\! \hat{e} \!\!&=&\!\! 
   \frac{ \!\!\left[ \hat{e}_{2} -  \frac{\hat{q}}{2}  - \frac{3\tilde{c}_{0}}{2} \bra{M}  \left\{ (F^{03})^2 - (F^{31})^2  \right\}  \ket{M} \right] }{-q^-} \nonumber \\
      &=& \!\!\! \frac{-1}{q^-}\! \left[ \hat{e}_{2} -  \frac{\hat{q}}{2}  + \frac{\tilde{c}_{0}}{4} \bra{M} \left[ F^{\mu\nu} F_{\mu\nu}   \right]  \ket{M} \right] \nonumber \\
       &=& \!\!\! \frac{-1}{q^-} \!  \left[ \hat{e}_{2} -  \frac{\hat{q}}{2}  + \frac{d_{0}}{T} \!\bra{M} \! \alpha_s \! \left[ F^{\mu\nu} F_{\mu\nu} \right] \! \ket{M} \right]\!\!, 
 \label{eq:EinsteinRelation}
\end{eqnarray}
where, $d_{0} = \sfrac{4 \pi^2}{(\delta_{T} \sqrt{2} N_c)}$. In the above equation, the repeated indices $\mu,\nu$ are summed over, from $0$-$3$, and the field strength product includes an implied trace.
The derived relation in Eq.~(\ref{eq:EinsteinRelation}) represents the fluctuation-dissipation relation for a hard quark traversing a non-perturbative thermalized gluon plasma. 
Our analysis indicates that the longitudinal drag $\hat{e}$ experienced by a hard quark is controlled by the imbalance between the longitudinal diffusion coefficient $\hat{e}_{2}$, the transverse momentum broadening coefficient $\hat{q}$, and the vacuum-subtracted thermal gluon condensate. Compared to the other coefficients, the light-cone drag coefficient $\hat{e}$ is suppressed by the inverse light-cone momentum of the hard quark, $q^-$.
While the derivation is carried out for a hard quark, it could very easily be extrapolated to the case of a hard gluon. The only change will be the color factors in the coefficient $d_0$. The fluctuation-dissipation relation derived here is in the limit that the hard quark has a single interaction with the medium, which makes a small change in the momenta of the quark. This relation should be distinguished from that derived in Ref.~\cite{Ghiglieri:2015ala}, which describes a hard parton about to thermalize within the plasma.
 
  In the high energy ($q^{-} \gg T$) limit, the drag coefficient $\hat{e}$ is determined by the vacuum subtracted thermal gluon condensate. In this regime, the longitudinal diffusion coefficient $\hat{e}_{2}$ is a sum of $\hat{q}/2$ and a constant times the thermal gluon condensate. Furthermore, the jet transport coefficient $\hat{q}$, given in Eq.~(\ref{eq:k2_perp_fnl}), can be expressed as the product of the renormalized coupling constant and the entropy density of the thermalized gluonic plasma~\cite{Kumar:2020wvb}.
In this effort, we had defined $\hat{e}_2 = \langle (k^-)^2 \rangle $, the resulting fluctuation dissipation relation can be expressed as $\hat{e}=-C/q^-$ [$C$ represents the term in brackets in Eq.~\eqref{eq:EinsteinRelation}]. If we had instead used the definition $\hat{e}_2 = \langle (k^-)^2 \rangle  - \langle k^- \rangle^2 = \langle (k^-)^2 \rangle - \hat{e}^2$, we would obtain a quadratic equation for $\hat{e}$: $ \hat{e}^2/q^- \!\!+\hat{e} + C/q^- = 0$. In the limit of large $q^-$, $\hat{e}$ would again admit a series solution $ \hat{e} = - C/q^- - C^2/(q^-)^3 +. \ldots$ Dropping terms suppressed by $(q^-)^3$ and higher, we obtain the same relation as Eq.~\eqref{eq:EinsteinRelation}.
As a result, at temperatures close to $T_C$,   $\hat{e}$ receives considerable contributions from the gluon condensate. This would indicate \emph{an enhancement of drag loss ($\hat{e}$) and longitudinal fluctuation ($\hat{e}_2$) near $T_C$}.



The calculation in this Letter is carried out by treating the interaction between the hard quark and the medium perturbatively, while plasma dynamics, including the gluon field-strength correlators, and their derivatives, are treated non-perturbatively. The energy scale of the propagating quark is assumed to be much larger than the temperature scale of the medium. The above relation was derived at leading order in the interaction of the hard quark with the medium. Eq.~\eqref{eq:EinsteinRelation} will receive modifications from both higher-order corrections to the perturbative interaction and from renormalization of the coupling constant and operator products~\cite{Blaizot:2014bha,Liou:2013qya,Iancu:2014kga,Kumar:2019uvu,Kumar:2025rsa}. While these corrections may introduce modifications in some of the numerical factors in Eq.~\eqref{eq:EinsteinRelation}, the qualitative form of the equation will remain unchanged.

Calculations in this paper were restricted to a gluonic (quenched) plasma. Extending this calculation to a Quark Gluon Plasma will also introduce flavor changing transport coefficients~\cite{Sirimanna:2022zje,Kumar:2025asj} which may introduce new relations. These will change Eq.~\eqref{eq:EinsteinRelation}, both by possibly introducing new transport coefficients and via mixing with the renormalized form of existing coefficients.

{\it Acknowledgments}:
A.~K. was supported in part by the Canada Research Chair [grant number CRC-2022-00146], the Natural Sciences and Engineering Research Council (NSERC) of Canada [grant number SAPIN-2023-00029], and the US National Science Foundation (NSF) [grant number OAC-2004571]. A.~M. was supported by the US Department of Energy (DOE) under grant number DE-SC0013460. 
I.~S. and J.~H.~W. were supported by the Deutsche Forschungsgemeinschaft (DFG) through the CRC-TR 211 ``Strong-interaction matter under extreme conditions'' (Project No. 315477589 - TRR 211). 


\bibliography{refs}

\end{document}